\documentclass[twocolumn,preprintnumbers,amsmath,amssymb]{revtex4}

\usepackage{graphicx}% Include figure files
\usepackage{dcolumn}% Align table columns on decimal point
\usepackage{bm}% bold math
\usepackage{subfigure}

\DeclareMathOperator{\erf}{erf}
\DeclareMathOperator{\erfgau}{erfgau}
\newcommand{\Psimu}{\ensuremath{\Psi^{\mu}}}
\newcommand{\mut}{\ensuremath{\tilde{\mu}}}
\newcommand{\bra}[1]{\ensuremath{\langle #1 \vert}}
\newcommand{\ket}[1]{\ensuremath{\vert #1  \rangle}}

\renewcommand{\b}[1]{\ensuremath{\mathbf{#1}}}

\begin{document}

%\preprint{APS/123-QED}

\title{Short-range exchange and correlation energy density functionals: beyond the local density approximation}

\author{Julien Toulouse}
\author{Fran\c{c}ois Colonna}
\author{Andreas Savin}
 \email{savin@lct.jussieu.fr}
\affiliation{
Laboratoire de Chimie Th\'eorique, CNRS et Universit\'e Pierre et Marie Curie,\\
4 place Jussieu, 75252 Paris, France
}

\date{\today}

\begin{abstract}
We propose approximations which go beyond the local density approximation for the short-range exchange and correlation density functionals appearing in a multi-determinantal extension of the Kohn-Sham scheme. A first approximation consists in defining locally the range of the interaction in the correlation functional. Another approximation, more conventional, is based on a gradient expansion of the short-range exchange-correlation functional. Finally, we also test a short-range generalized-gradient approximation by extending the Perdew-Burke-Ernzerhof exchange-correlation functional to short-range interactions.
\end{abstract}

\maketitle

\section{Introduction}
\label{sec:intro}

The Kohn-Sham (KS) scheme~\cite{KohSha-PR-65} of density functional theory (DFT) (see, e.g.,~\cite{ParYan-BOOK-89,DreGro-BOOK-90}) can be extended to handle multi-determinantal wave functions~\cite{SavColPol-IJQC-03}. This enables to describe near-degeneracy correlation effects by explicitly taking into account several Slater determinants. The method is based on a decomposition of the true Coulomb electron-electron interaction as
\begin{equation}
\label{}
\frac{1}{r} = v_{ee}^{\mu}(r) + \bar{v}_{ee}^{\mu}(r),
\end{equation}
where $v_{ee}^{\mu}(r)$ is a long-range interaction and $\bar{v}_{ee}^{\mu}(r)$
is the complement short-range interaction. This separation is controlled by the
parameter $\mu$. In previous studies~\cite{Sav-INC-96,LeiStoWerSav-CPL-97,PolSavLeiSto-JCP-02}, the error function has been used to define the long-range interaction
\begin{equation}
\label{veeerf}
v_{ee,\erf}^{\mu}(r)=\frac{\erf(\mu r)}{r},
\end{equation}
referred to as the \textit{erf} interaction. More recently~\cite{TouColSav-JJJ-XX}, we have proposed a sharper long-range/short-range separation with the \textit{erfgau} interaction
\begin{equation}
\label{veeerfgau}
v_{ee,\erfgau}^{\mu}(r)=\frac{\erf(c \mu r)}{r} - \frac{2 c \mu}{\sqrt{\pi}} e^{-\frac{1}{3}c^2 \mu^2 r^2},
\end{equation}
where $c = \left( 1+6\sqrt{3}\right)^{1/2} \approx 3.375$ is a constant chosen to facilitate the comparison with the \textit{erf} interaction. Indeed, with this choice, the parameter $\mu$ has roughly the same physical meaning for the \textit{erf} and \textit{erfgau} interactions: $1/\mu$ represents the distance beyond which the interaction reduces to the usual Coulomb long-range tail. (Note that in Ref.~\onlinecite{TouColSav-JJJ-XX} the constant $c$ was absorbed into the parameter $\mu$). Both modified interactions define a generalized adiabatic connection~\cite{Yan-JCP-98} between the non-interacting KS system corresponding to $\mu=0$ and the physical interacting system corresponding to $\mu \to \infty$.

The ground-state energy of an electronic system can then be (exactly in principle) calculated by the sum of a wave function part associated to the long-range electron-electron interaction $\hat{V}_{ee}^{\mu}=\sum_{i<j}v_{ee}^{\mu}(r_{ij})$ and a density functional part describing the remaining interactions
\begin{equation}
E =  \bra{\Psimu} \hat{T}+\hat{V}_{ee}^{\mu}+\hat{V}_{ne} \ket{\Psimu} + \bar{U}^{\mu}[n] + \bar{E}^{\mu}_{xc}[n],
\label{E}
\end{equation}
where $\hat{T}$ is the kinetic energy operator, $\hat{V}_{ne}=\sum_{i}v_{ne}(\b{r}_i)$ is the nuclei-electron interaction, $\bar{U}^{\mu}$ is the short-range Hartree energy and $\bar{E}^{\mu}_{xc}$ is the short-range exchange-correlation functional defined as the difference between the standard KS exchange-correlation energy $E_{xc}$ and the long-range exchange-correlation energy $E^{\mu}_{xc}$ associated to the interaction $v_{ee}^{\mu}$
\begin{equation}
\label{}
\bar{E}^{\mu}_{xc} = E_{xc} - E^{\mu}_{xc}.
\end{equation}
In Eq.~(\ref{E}), $\Psimu$ is the multi-determinantal ground-state wave function of a fictitious system with long-range electron-electron interaction $\hat{V}_{ee}^{\mu}$ 
\begin{equation}
\left( \hat{T}+\hat{V}_{ee}^{\mu}+\hat{V}^{\mu} \right) \ket{\Psimu} = E^{\mu} \ket{\Psimu},
\label{HmuPsimu}
\end{equation}
where $\hat{V}^{\mu}=\sum_{i} v^{\mu}(\b{r}_i)$ is the external local potential ensuring that this fictitious system has the same ground state density $n$ than the physical system. The potential $v^{\mu}$ is simply related to the functional derivative of the short-range functionals through
\begin{equation}
v^{\mu}(\b{r})=v_{ne}(\b{r})+ \frac{ \delta\bar{U}^{\mu}}{\delta n(\b{r})} +\frac{\delta \bar{E}^{\mu}_{xc}}{\delta n(\b{r})}.
\label{vmu}
\end{equation}

Previous applications of the method~\cite{LeiStoWerSav-CPL-97,PolSavLeiSto-JCP-02} show that, for a reasonable long-range/short-range separation ($\mu$ not too small) and if the few most important (nearly degenerate) configurations are included in the wave function $\Psimu$, good results are obtained for total atomic and molecular energies, including near-degenerate systems. This is remarkable since, in these previous studies, only the simple local density approximation (LDA) was used for the short-range functional $\bar{E}^{\mu}_{xc}[n]$. 

The purpose of the present work is to further improve the accuracy of the method by proposing approximations for $\bar{E}^{\mu}_{xc}[n]$ which go beyond the LDA. We will not present in this paper total energies $E$ (or equivalently total exchange-correlation energies $E_{xc}$), but will focus only on short-range exchange-correlation energies $\bar{E}^{\mu}_{xc}$. Indeed, for a chosen $\mu$, it is the approximation for $\bar{E}^{\mu}_{xc}[n]$ that limits in practice the accuracy of the method, not the other long-range contributions to the energy that can be calculated by a traditional \textit{ab initio} wave function method. We will test our proposed approximations for $\bar{E}^{\mu}_{xc}$ by comparison to accurate data obtained for small atomic systems.

The paper is organized  as follows. In Sec.~\ref{sec:limitLDA}, we discuss the limitations of the LDA for $\bar{E}^{\mu}_{xc}[n]$. In Sec.~\ref{sec:localmu}, we introduce a local interaction parameter $\mu$ to improve the LDA correlation functional. We then discuss the extension of traditional gradient corrections to the short-range functional $\bar{E}^{\mu}_{xc}[n]$. Indeed, the short-range versions of the gradient expansion approximation (GEA) is presented in Sec.~\ref{sec:gea}, while Sec.~\ref{sec:pbe} contains the extension of a generalized-gradient approximation (GGA), namely the PBE functional~\cite{PerBurErn-PRL-96}, to a modified interaction. Finally, Sec.~\ref{sec:conclusion} contains our concluding remarks.

Atomic units will be used throughout this work.

\section{Limitations of the LDA}
\label{sec:limitLDA}
In the LDA, the short-range exchange-correlation energy per particle $\bar{\varepsilon}_{xc}^{\mu,\text{unif}}(n)$ of a uniform electron gas with modified interaction~\cite{Sav-INC-96,TouSavFla-IJQC-XX} is transferred locally to the inhomogeneous system of interest
\begin{equation}
\label{}
\bar{E}^{\mu}_{xc}[n]= \int n(\b{r}) \, \bar{\varepsilon}^{\mu,\text{unif}}_{xc}(n(\b{r})) d\b{r}.
\end{equation}
To underline the dependence on the interaction parameter, we will refer to this approximation as $\mu$-LDA. 

\begin{figure}
\includegraphics[scale=0.75]{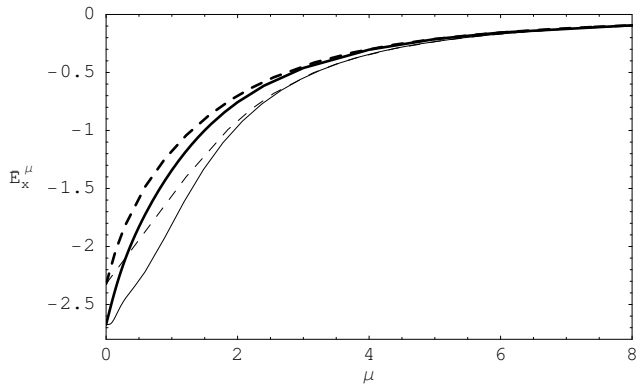}
\caption{Accurate (full curves) and $\mu$-LDA (dashed curves) short-range exchange energy along the \textit{erf} (thick curves) and \textit{erfgau} (thin curves) adiabatic connections for the Be atom.
}
\label{fig:ex-be-erferfgau-lda}
\end{figure}

\begin{figure}
\includegraphics[scale=0.75]{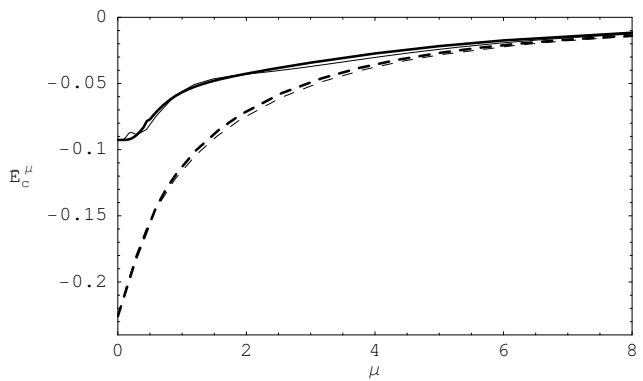}
\caption{Accurate (full curves) and $\mu$-LDA (dashed curves) short-range correlation energy along the \textit{erf} (thick curves) and \textit{erfgau} (thin curves) adiabatic connections for the Be atom.
}
\label{fig:ec-be-erferfgau-lda}
\end{figure}

In Figs.~\ref{fig:ex-be-erferfgau-lda} and~\ref{fig:ec-be-erferfgau-lda}, we have plotted the short-range LDA exchange and correlation energies of the Be atom with respect to $\mu$ for the \textit{erf} and \textit{erfgau} interactions. An accurate density obtained from a multi-reference configuration interaction calculation with single and double excitations (MRCISD)~\cite{WerKno-JCP-88,KnoWer-CPL-88} has been used. For comparison, accurate calculations of the exchange and correlations energies along the adiabatic connections are also reported. In these accurate calculations, we start from an accurate reference density and, for each $\mu$, numerically optimize the external potential $v^{\mu}(\b{r})$ appearing in Eq.~(\ref{HmuPsimu}) so as to recover the reference density. The ground-state wave function $\Psimu$ is then computed according to Eq.~(\ref{HmuPsimu}) by MRCISD and the various energy components like the short-range exchange and correlation energies are deduced. (For further details, see Refs.~\onlinecite{ColSav-JCP-99,PolColLeiStoWerSav-IJQC-03,TouColSav-JJJ-XXa}). 

For both the \textit{erf} and \textit{erfgau} interactions, the $\mu$-LDA is very accurate for large $\mu$ but fails near the KS end ($\mu=0$) of the adiabatic connection. In particular, the exchange energy is underestimated and the correlation energy is overestimated. Thanks to a sharper separation of long-range and short-range electron-electron interactions, the \textit{erfgau} interaction provides a slight improvement over the \textit{erf} interaction. In fact, Fig.~\ref{fig:ex-be-erferfgau-lda} shows that the \textit{erfgau} $\mu$-LDA exchange energy curve reaches the exact one for a larger energy ($\approx -0.75$ Hartree) and a smaller $\mu$ ($\approx 2.5$) than in the \textit{erf} case ($\approx -0.5$ Hartree at $\mu \approx 3$), meaning that the \textit{erfgau} $\mu$-LDA is able to correctly describe a larger part of the exchange energy. For the correlation energy, the two interactions lead to similar results. We note in passing that the unimportant little bump on the curve of the accurate correlation energy for small $\mu$ with the \textit{erfgau} interaction in Fig.~\ref{fig:ec-be-erferfgau-lda} (and in other following figures) is a manifestation of the non-monotonicity of this interaction with respect to $\mu$ (being in turn responsible for its attractive character for very small $\mu$~\cite{TouSavFla-IJQC-XX}). In the remaining of the paper, we will present results only for the \textit{erfgau} interaction.

\begin{figure*}
\includegraphics[scale =0.70]{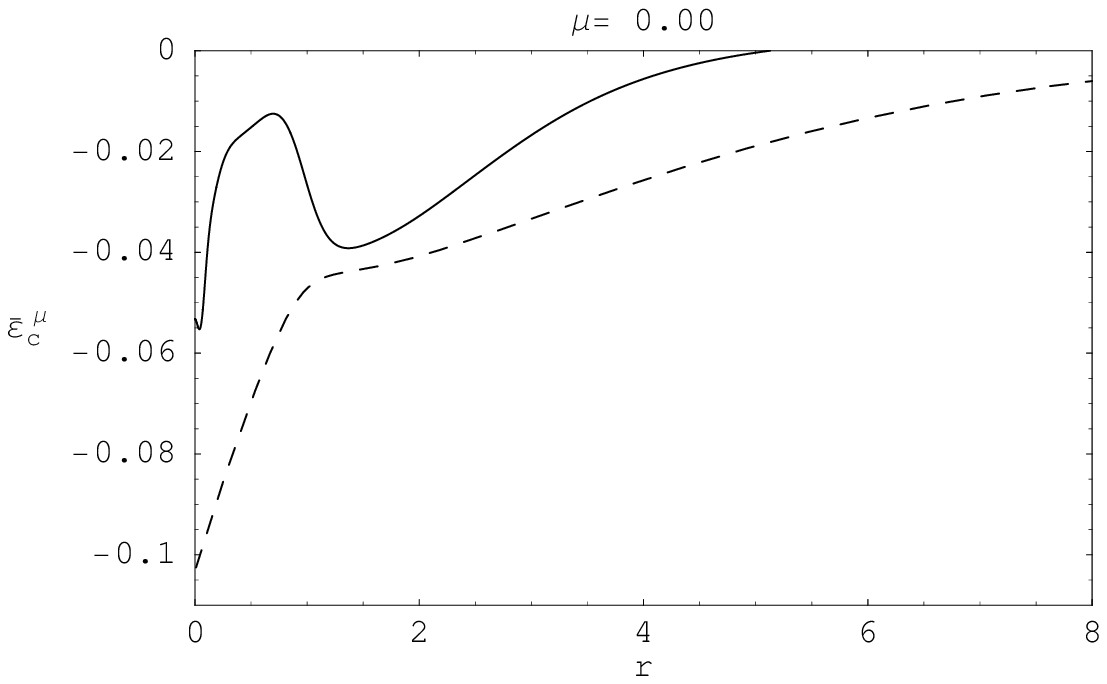}
\includegraphics[scale =0.70]{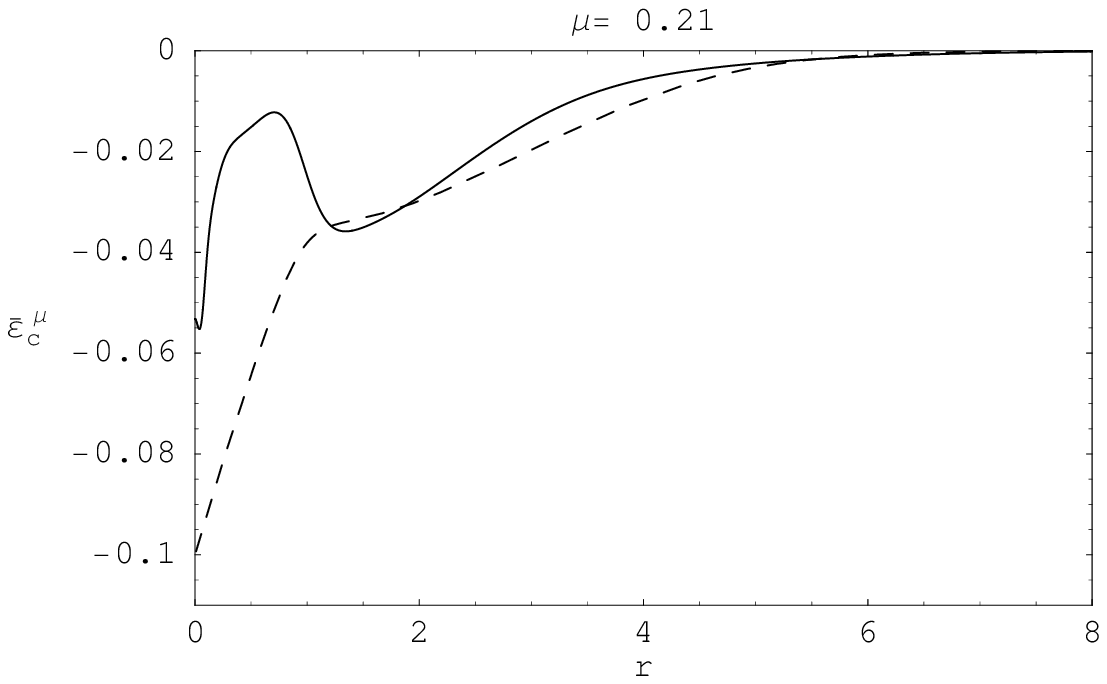}
\includegraphics[scale =0.70]{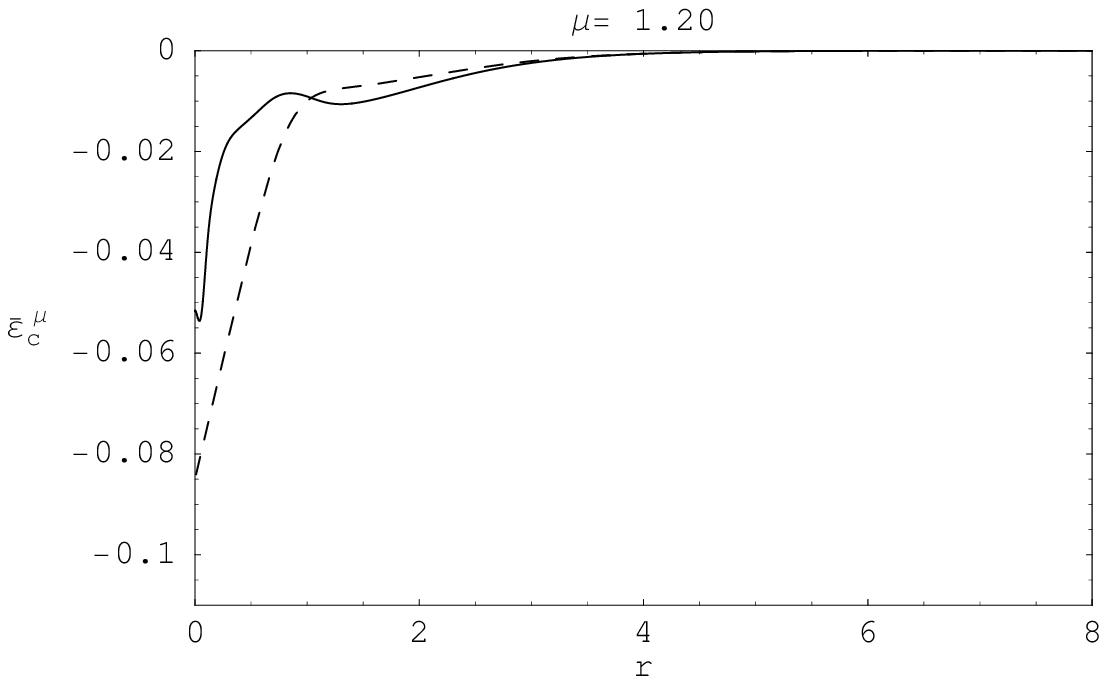}
\includegraphics[scale =0.70]{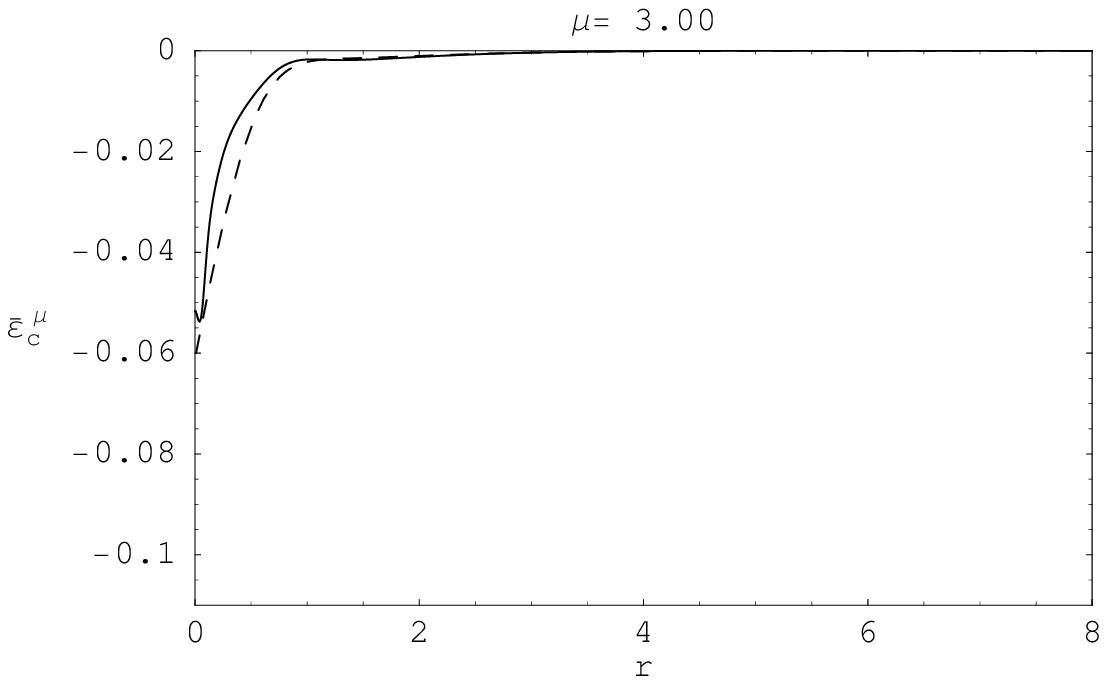}
\caption{Short-range correlation energy density $\bar{\varepsilon}_{c}^{\mu}(r)$ with respect to $r$ for the Be atom with the \textit{erfgau} interaction for $\mu=0$, $0.21$, $1.20$, $3.00$. Accurate data computed from integration over the adiabatic connection (Eq.~\ref{epsc}, full curve) are compared with the $\mu$-LDA (dashed curve).
}
\label{fig:epsc-be-erfgau-mu0.2-lda}
\end{figure*}

The performance of the $\mu$-LDA can be further analyzed by looking, for instance, at the correlation energy density $\bar{\varepsilon}_{c}^{\mu}(r)$. There is not unique definition of this quantity; we used the definition of Ref.~\onlinecite{ColMaySav-PRA-03} or~\onlinecite{TouColSav-JJJ-XXa} based on the integration over the adiabatic connection
\begin{equation}
\bar{\varepsilon}_{c}^{\mu}(\b{r}_1) = \frac{1}{2}  \int_{\mu}^{\infty} d\xi \int d\b{r}_2 n_{c}^{\xi}(\b{r}_1,\b{r}_2) \frac{\partial v_{ee}^{\xi}(r_{12})}{\partial \xi},
\label{epsc}
\end{equation}
where $n_{c}^{\xi}(\b{r}_1,\b{r}_2)$ is the correlation hole for the interaction parameter $\xi$. In Fig.~\ref{fig:epsc-be-erfgau-mu0.2-lda}, this correlation energy density accurately computed for the Be atom is compared with the $\mu$-LDA with the \textit{erfgau} interaction for a series of $\mu$'s. For $\mu=0$ (KS system), the correlation energy density of Eq.~(\ref{epsc}) is largely overestimated by the LDA over the whole range of $r$. When $\mu$ is increased (Fig.~\ref{fig:epsc-be-erfgau-mu0.2-lda} with $\mu=0.21$ and $\mu=1.20$), the $\mu$-LDA starts to better reproduces the accurate energy density in the valence region ($r \gtrsim 1$) but still overestimates it in the core region ($r \lesssim 1$). For the $\mu$-LDA to also well reproduce the core region, larger values of $\mu$ are required (Fig.~\ref{fig:epsc-be-erfgau-mu0.2-lda} with $\mu=3.00$).

Having in mind that the choice of a large value of $\mu$ means an important computational effort for the part of the energy treated by wave function methods, it is important to improve the short-range exchange-correlation functional toward the small $\mu$ region of the adiabatic connection.

\section{LDA correlation with local interaction parameter}
\label{sec:localmu}

In inhomogeneous systems the electrons are correlated on a shorter distance than in the uniform electron gas. Thus, within the KS scheme, the LDA transfers spurious long-range correlation contributions from the electron gas to the finite system of interest. This point have long been understood and have guided important developments in density functional approximations. For instance, Langreth and Mehl~\cite{LanMeh-PRL-81,LanMeh-PRB-83} proposed a gradient correction to the LDA based on a cut-off in momentum-space for small $k$ which basically corresponds to removing the long-range interaction effects.

In the method proposed here, for small $\mu$, i.e. when a large range of the electron-electron interaction is retained in the functional part, the $\mu$-LDA also transfers spurious long-range interaction effects from the uniform electron gas to finite systems and leads to an overestimation of the correlation energy. A simple way to eliminate this incorrect long-range interaction effects, is to use in the $\mu$-LDA correlation functional an effective interaction parameter, larger than the one used in the wave function part of the calculation. Of course, in the uniform electron gas, the range of the interaction is relative to the density (the relevant scaled variable is $r/r_s$ with $r_s=(4\pi n/3)^{-1/3}$) and consequently one has to actually consider an effective local interaction parameter, $\mu_l(\b{r})$, defined from the inhomogeneous density profile $n(\b{r})$. However, the interaction parameter used in the functional must not be smaller that the global $\mu$ used in the wave function part of the calculation to avoid a double counting of correlation effects. We therefore take in the correlation $\mu$-LDA functional an effective local interaction parameter given by
\begin{equation}
\label{}
\mu_{\text{eff}}(\b{r})=\max(\mu_l(\b{r}),\mu).
\end{equation}
Thus, this approximation corresponds to simply drop off some long-range correlation from the LDA in regions of space where $\mu_l(\b{r}) > \mu$.

Several choices are possible for the local interaction parameter $\mu_l(\b{r})$. The first natural length scale in real space for the electron-electron interaction is provided by the Wigner-Seitz radius $r_s$, representing the radius of the sphere centered in $\b{r}$ containing one electron. Equivalently, the same length scale can be expressed in momentum space by the Fermi wave vector $k_F=1/(\alpha r_s)$ where $\alpha=(9\pi/4)^{-1/3}$. Following a previous work~\cite{PolSavLeiSto-JCP-02}, we thus take as a local interaction parameter
\begin{equation}
\mu_{l,\text{a}}(\b{r})= \frac{1}{r_s(\b{r})} = \alpha k_F(\b{r}).
\label{mula}
\end{equation}
However, studies of the uniform electron gas indicate that the relevant length scale for correlation effects in momentum space is given by the Thomas-Fermi wave vector $k_s=\sqrt{4 k_F/\pi}$ describing the screening of the Coulomb interaction. Similarly to Eq.~(\ref{mula}), we can thus take as local interaction parameter
\begin{equation}
\mu_{l,\text{b}}(\b{r})=\alpha k_s(\b{r})=\frac{2\sqrt{\alpha/\pi}}{\sqrt{r_s(\b{r})}}.
\label{mulb}
\end{equation}
More elaborated dependences of $\mu_{l}$ on $r_s$ are possible. For example, Ferrell~\cite{Fer-PR-57} has determined a momentum cut-off in the uniform electron gas corresponding to the shortest wavelength at which plasma oscillations can be sustained by the electron gas. It turns out however that results obtained with a local interaction parameter set to Ferrel's momentum cut-off are very similar to those obtained with the local interaction parameter of Eq.~(\ref{mulb}) and will not be discussed here. 

A much more interesting possibility is to choose a local interaction parameter depending on the gradient of the density. In fact, as summarized by Svendsen and von Barth~\cite{SveBar-IJQC-95} from the work of Langreth and Mehl~\cite{LanMeh-PRL-81,LanMeh-PRB-83} and Perdew \textit{et al.}~\cite{PerCheVosJacPedSinFio-PRB-92}, ``distances over which electrons are correlated in realistic inhomogeneous systems are determined more by actual variations in the density profile than by the screening of an electron gas at the local value of the density''. We use a simple geometrical argument similar of that of Langreth and Mehl~\cite{LanMeh-PRL-81} to determine a local characteristic distance $d(\b{r})$ over which the density varies in an inhomogeneous system. Consider a density distribution with constant density gradient consisting of an isosceles triangle as a prototype of an atom. The extension of this density distribution is $d=4 n /|\nabla n|$ where $n$ is the density at the middle of one side of the triangle, i.e. a typical value of the density. We thus define a (semi)local interaction parameter by
\begin{equation}
\mu_{l,\text{c}}(\b{r})= \frac{|\nabla n(\b{r})|}{ 4 n(\b{r})}.
\label{mulc}
\end{equation}
The constant $4$ in Eq.~(\ref{mulc}) is somehow arbitrary; experience shows that this is reasonable.

\begin{figure}
\includegraphics[scale=0.75]{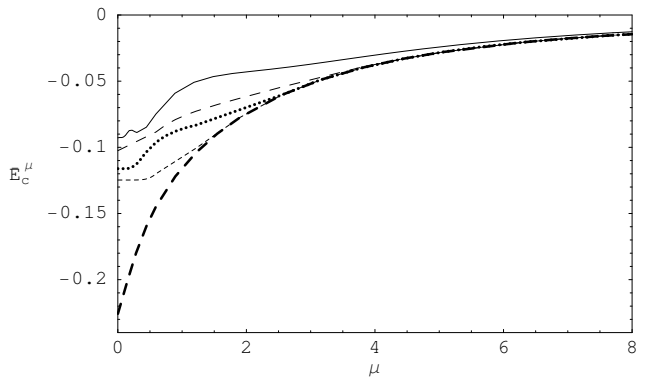}
\caption{Short-range correlation energy along the \textit{erfgau} adiabatic connection for the Be atom. Accurate data (full curve) are compared to the $\mu$-LDA functional with a global $\mu$ (thick long-dashed curve) and with (semi)local interaction parameters: $\mu_{l,\text{a}}=\alpha k_F$ (long-dashed curve), $\mu_{l,\text{b}}=\alpha k_s$ (short-dashed curve) and $\mu_{l,\text{c}}= |\nabla n|/(4 n)$ (dotted curve).
}
\label{fig:ec-be-erfgau-localmu}
\end{figure}

\begin{figure}
\includegraphics[scale=0.75]{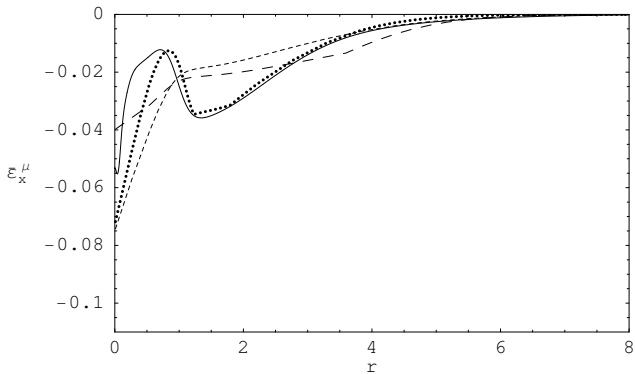}
\caption{Short-range correlation energy density $\bar{\varepsilon}_{c}^{\mu}(r)$ with respect to $r$ for the Be atom with the \textit{erfgau} interaction for $\mu=0.21$. Accurate data (Eq.~\ref{epsc}, full curve) computed from integration over the adiabatic connection are compared with the $\mu$-LDA functional with local interaction parameters: $\mu_{l,\text{a}}=\alpha k_F$ (long-dashed curve), $\mu_{l,\text{b}}=\alpha k_s$ (short-dashed curve) and $\mu_{l,\text{c}}= |\nabla n|/(4 n)$ (dotted curve).
}
\label{fig:epsc-be-erfgau-mu0.2-localmu}
\end{figure}

Fig.~\ref{fig:ec-be-erfgau-localmu} shows the correlation energy is greatly improved by the local $\mu$ approach.  However, for the choices $\mu_{l,\text{a}}=\alpha k_F$ or $\mu_{l,\text{b}}=\alpha k_s$, a more detailed analysis through the correlation energy density in Fig.~\ref{fig:epsc-be-erfgau-mu0.2-localmu} indicates the $\mu$-LDA functional with a local $\mu$ is only an average of the ``exact'' correlation energy density. Of course, the $\mu$-LDA functional with a local $\mu$ improves on increasing of $\mu$ (not shown) but it is clear that both choices are inadequate for describing the shell structure. On the contrary, one sees on the same Fig.~\ref{fig:epsc-be-erfgau-mu0.2-localmu} that the choice $\mu_{l,\text{c}}= |\nabla n|/(4 n)$ enables to recover well the shell structure in Be. We therefore consider this last choice as the more appropriate local parameter $\mu_l$ for improving the LDA correlation functional.

Finally, we note that this local $\mu$ approach cannot be directly applied to the exchange energy. Indeed, the LDA underestimates the exchange energy and thus choosing a large effective interaction parameter in the $\mu$-LDA would only deteriorate the exchange energy. 

\section{Gradient expansions}
\label{sec:gea}

We now consider more conventional approaches to correct the local density approximation at small interaction parameter, namely gradient corrections.
In the gradient expansion approximation (GEA), the short-range exchange energy density is expanded in term of the reduced gradient $s=|\nabla n|/(2 k_F n)$ and the reduced Laplacian $q=\nabla^2 n/((2 k_F)^2 n)$
\begin{equation}
\bar{\varepsilon}_{x}^{\mu,\text{GEA-Lap}}(n) = \bar{\varepsilon}_{x}^{\mu,\text{unif}}(n) \left[1+ a(\mut) s^2 + d(\mut) q \right],
\label{epsxGEA-Lap}
\end{equation}
where the reduced interaction parameter $\mut=\mu/(2 k_F)$ has been introduced. The gradient and Laplacian coefficients $a(\mut)$ and $d(\mut)$ can be calculated numerically for all $\mu$'s except for $\mu=0$ where there are not defined (see Appendix~\ref{app:geax}). Alternatively, one can perform an integration by parts and obtain the more conventional GEA exchange density
\begin{equation}
\bar{\varepsilon}_{x}^{\mu,\text{GEA}}(n) = \bar{\varepsilon}_{x}^{\mu,\text{unif}}(n) \left[1+ b(\mut) s^2 \right],
\label{epsxGEA}
\end{equation}
where the exchange second-order gradient coefficient $b(\mut)$ has been calculated analytically for the \textit{erf} and \textit{erfgau} interactions [see Appendix~\ref{app:geax}, Eqs.~(\ref{berf}) and~(\ref{berfgau})]. The approximations of Eqs.~(\ref{epsxGEA-Lap}) and~(\ref{epsxGEA}) will be referred to as $\mu$-GEA-Lap and $\mu$-GEA, respectively.

\begin{figure}
\includegraphics[scale=0.75]{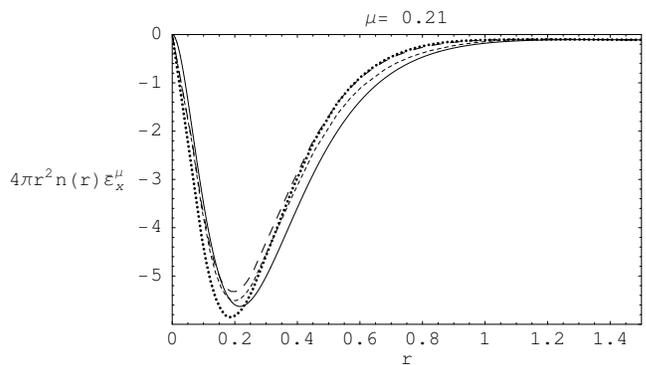}
\caption{Radial short-range exchange energy density $4\pi r^2 n(r) \bar{\varepsilon}_{x}^{\mu}(r)$ with respect to $r$ for the Be atom with the \textit{erfgau} interaction for $\mu=0.21$. Accurate data (Eq.~\ref{epsx}, full curve) are compared with the $\mu$-LDA (long-dashed curve), the $\mu$-GEA (short-dashed curve) and the $\mu$-GEA-Lap (dotted curve).
}
\label{fig:epsx-be-erfgau-mu0.21-gea}
\end{figure}

Fig.~\ref{fig:epsx-be-erfgau-mu0.21-gea} compares the $\mu$-GEA-Lap and $\mu$-GEA short-range exchange energy densities for the Be atom with the \textit{erfgau} interaction at $\mu=0.21$. The exchange energy density in the $\mu$-LDA and calculated from the accurate exchange hole $n_{x}(\b{r}_1,\b{r}_2)$,
\begin{equation}
\bar{\varepsilon}_{x}^{\mu}(\b{r}_1) = \frac{1}{2} \int d\b{r}_2 n_{x}(\b{r}_1,\b{r}_2) \bar{v}_{ee}^{\mu}(r_{12}),
\label{epsx}
\end{equation}
are also reported. To avoid the Laplacian divergence at the nucleus in $\mu$-GEA-Lap, radial energy densities $4\pi r^2 n(r) \bar{\varepsilon}_{x}^{\mu}(r)$ are actually represented. One sees that the integration by parts in $\mu$-GEA does not change qualitatively the shape of the curve and it is meaningful to compare both the $\mu$-GEA-Lap and $\mu$-GEA exchange energy densities to the accurate and $\mu$-LDA ones. Besides, for larger values of $\mu$, $\mu$-GEA-Lap and $\mu$-GEA become nearly identical (not shown). Of course, $\mu$-GEA-Lap and $\mu$-GEA give the same short-range exchange energy $\bar{E}_{x}^{\mu}$ and we will used in practice $\mu$-GEA because of its greatest simplicity.

Similar to Eq.~(\ref{epsxGEA}), the short-range correlation energy density is expanded in term of the reduced gradient $t=|\nabla n|/(2 k_s n)$
\begin{equation}
\label{}
\bar{\varepsilon}_{c}^{\mu,\text{GEA}}(n) = \bar{\varepsilon}_{c}^{\mu,\text{unif}}(n)  + \beta (r_s,\mu) t^2.
\end{equation}
The correlation second-order gradient coefficient $\beta (r_s,\mu)$ can be estimated by imposing that the correlation gradient expansion cancels the exchange gradient expansion
\begin{equation}
\label{betamu}
\beta(r_s,\mu) = -\bar{\varepsilon}_{x}^{\mu,\text{unif}}(n) b(\mut) (k_s/k_F)^2.
\end{equation}
This is based on the observation that, for $\mu=0$, the exchange-correlation contribution to the linear response of the uniform electron gas to a slowly oscillating perturbation (small momenta) almost vanishes~\cite{MorCepSen-PRL-95}. This short-range GEA functional approximation will be referred to as $\mu$-GEA. Our interest is to evaluate the improvement brought by $\mu$-GEA over $\mu$-LDA for the exchange and correlation energies separately. Notice that, in our construction, if we consider exchange and correlation together $\mu$-GEA is identical to $\mu$-LDA.

\begin{figure}
\includegraphics[scale=0.75]{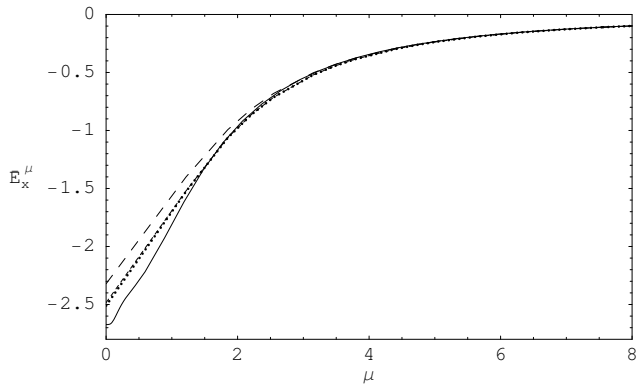}
\caption{Short-range exchange energy along the \textit{erfgau} adiabatic connection for the Be atom. Accurate data (full curve) are compared to the $\mu$-LDA functional (long-dashed curve), the $\mu$-GEA functional (dotted curve) and the $\mu$-PBE functional (short-dashed curve). The curves corresponding to the $\mu$-GEA and $\mu$-PBE functionals are nearly superimposed.
}
\label{fig:ex-be-erfgau-geapbe}
\end{figure}

\begin{figure}
\includegraphics[scale=0.75]{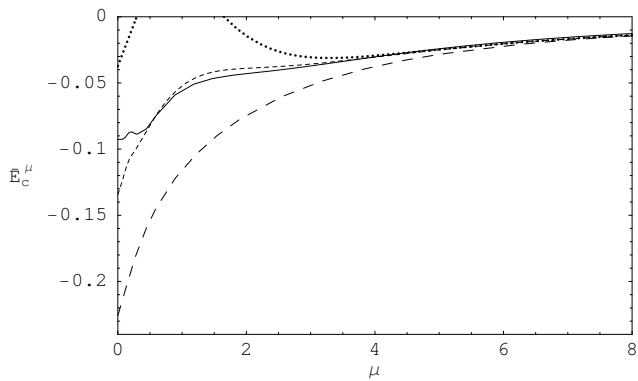}
\caption{Short-range correlation energy along the \textit{erfgau} adiabatic connection for the Be atom. Accurate data (full curve) are compared to the $\mu$-LDA functional (long-dashed curve), the $\mu$-GEA functional (dotted curve) and the $\mu$-PBE functional (short-dashed curve).
}
\label{fig:ec-be-erfgau-geapbe}
\end{figure}

The $\mu$-GEA exchange energy is plotted in Fig~\ref{fig:ex-be-erfgau-geapbe}. For large $\mu$ ($\mu \gtrsim 3$), the range of the interaction is too small to feel the slowly oscillating gradient correction and the $\mu$-GEA functional reduces to the $\mu$-LDA functional. The $\mu$-GEA curve reaches the exact one at a smaller value of $\mu$ ($\mu \approx 2$) than the $\mu$-LDA curve ($\mu \approx 3$). For small $\mu$, the $\mu$-GEA reduces the $\mu$-LDA error by about a factor two. For the correlation energy (Fig.~\ref{fig:ec-be-erfgau-geapbe}), the $\mu$-GEA also extends the domain of validity of the $\mu$-LDA when $\mu$ is decreased down to $\mu \approx 4$. For smaller $\mu$ when long-range correlation effects are introduced in the functional, the gradient expansion breaks down.

Whereas the GEA for the KS scheme ($\mu=0$) often disappointingly deteriorates the LDA in real inhomogeneous systems~\cite{MaBru-PR-68,PerLanSah-PRL-77}, it can be consider as a systematic way to extend the domain of validity of the $\mu$-LDA for the short-range part only of the exchange and correlation functionals. These results are in agreement with the wave vector analysis of the GEA by Langreth and Perdew~\cite{LanPer-PRB-80} who found that the GEA works well for large momenta (corresponding to short-range density fluctuations).

\section{Short-range PBE functional}
\label{sec:pbe}

Starting from the gradient expansions of the previous section, it is possible to construct generalized-gradient approximations (GGA) for the short-range exchange and correlation energy functionals which cure the main shortcomings of the $\mu$-GEA. We have actually constructed a modified version of the PBE functional~\cite{PerBurErn-PRL-96} for the modified interactions. All of the details of the construction of this functional that we will call $\mu$-PBE are given in Appendix~\ref{app:PBE}. We simply mention here that we have use the same ansatz than PBE and impose similar theoretical constraints. However, a difference with the PBE approach is that we impose the second-order gradient coefficient for exchange and approximate that of correlation from it while PBE did the opposite. Consequently, our $\mu$-PBE functional does not reduce to the original PBE for $\mu=0$.

\begin{figure}
\includegraphics[scale=0.75]{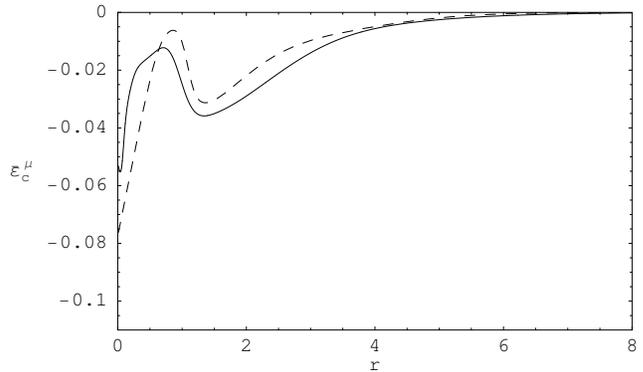}
\caption{Short-range correlation energy density $\bar{\varepsilon}_{c}^{\mu}(r)$ with respect to $r$ for the Be atom with the \textit{erfgau} interaction for $\mu=0.21$. Accurate data (full curve) computed from integration over the adiabatic connection are compared with the $\mu$-PBE functional (dashed curve).
}
\label{fig:epsc-be-erfgau-mu0.2-pbe}
\end{figure}

The exchange and correlation energies of the Be atom obtained with this $\mu$-PBE functional along the \textit{erfgau} adiabatic connection are compared to the $\mu$-LDA and $\mu$-GEA functionals in Figs.~\ref{fig:ex-be-erfgau-geapbe} and~\ref{fig:ec-be-erfgau-geapbe}. For a simple system like the Be atom, the $\mu$-GEA and $\mu$-PBE functionals are nearly identical for the exchange energy. For the correlation part, one sees that $\mu$-PBE greatly improved the $\mu$-LDA and $\mu$-GEA results. The $\mu$-PBE correlation energy is indeed very accurate along the entire adiabatic connection, except near the KS end ($\mu=0$) where a significant error remains. This inaccuracy of (semi)local functionals at $\mu=0$ is in favor of the strategy consisting in escaping the KS scheme by using a short-range functional at finite $\mu$ and treating the remaining part of the energy by other, more appropriate methods. In Fig.~\ref{fig:epsc-be-erfgau-mu0.2-pbe}, it is shown that the correlation energy density of $\mu$-PBE functional qualitatively describes the shell structure of the Be atom for small $\mu$.

\begin{figure}
\includegraphics[scale=0.75]{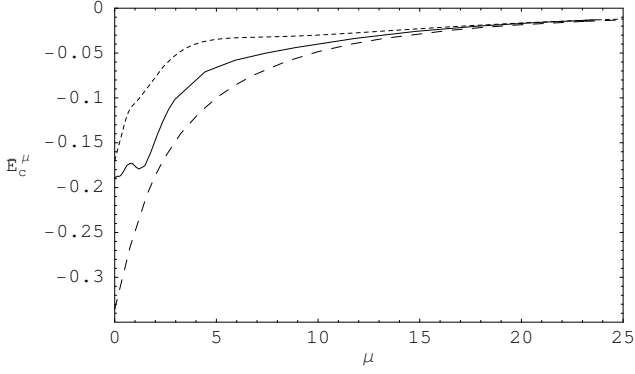}
\caption{Short-range correlation energy along the \textit{erfgau} adiabatic connection for the Ne$^{6+}$ atom. Accurate data (full curve) are compared to the $\mu$-LDA functional (long-dashed curve) and the $\mu$-PBE functional (short-dashed curve).
}
\label{fig:ec-ne6p-erfgau-pbe}
\end{figure}

Finally, the difficult case of the Ne$^{6+}$ atom, presenting strong near-degeneracy correlation effects, is reported in Fig.~\ref{fig:ec-ne6p-erfgau-pbe}. The correlation energy given by the $\mu$-PBE functional actually over-corrects that of the $\mu-$LDA. Again, this result suggests that for small $\mu$ semilocal approximations like the $\mu$-PBE functional are inappropriate, especially when near-degeneracy correlation effects play an important role. 

\section{Concluding remarks}
\label{sec:conclusion}

In this paper, we have considered exchange and correlation energy functionals associated to short-range electron-electron interactions. We have proposed new approximations which partially correct the local density approximation. Indeed, we showed that the short-range LDA correlation energy can be significantly corrected by simply adjusting in the functional the range of the interaction locally from the density and its gradient. We have also constructed generalizations of the GEA and of a GGA (namely, the PBE functional) to the case of short-range interactions. These new short-range functionals treat well a larger range of electron-electron interaction than the short-range LDA. However, better (possibly non-local) approximations are still needed to further extend the range of interaction well treated by the functional.

\appendix

\section{Gradient expansion of the short-range exchange energy functional}
\label{app:geax}
The short-range exchange energy functional writes
\begin{equation}
\label{Ex}
\bar{E}_{x}^{\mu} = \frac{1}{2} \iint n(\b{r}) n_x(\b{r},\b{r}+\b{R}) \bar{v}_{ee}^{\mu}(R) d\b{r} d\b{R},
\end{equation}
where $n_x(\b{r},\b{r}+\b{R})$ is the exchange hole. In the gradient-expansion approximation (GEA), the exchange hole is expanded to second order in term of the gradients of the density~\cite{GroDre-ZPA-81,PerWan-PRB-86}
\begin{equation}
\label{}
n_x^{\text{GEA}}(\b{r},\b{r}+\b{R}) = -\frac{1}{2} n(\b{r}) y(\b{r},\b{r}+\b{R}),
\end{equation}
where
\begin{eqnarray}
y &=& J + L k_{F}^{-3} \hat{\b{R}} \cdot \nabla k_{F}^{2} + M k_{F}^{-6} (\hat{\b{R}} \cdot \nabla k_{F}^{2})^2 
\nonumber\\
&& + N k_{F}^{-6} (\nabla k_{F}^{2})^2
 + z L k_{F}^{-4} (\hat{\b{R}} \cdot \nabla)^2 k_{F}^{2}/6 
\nonumber\\
&& -z^2 J k_{F}^{-4} \nabla^2 k_{F}^{2}/48,
\label{}
\end{eqnarray}
with $k_F(\b{r})=(3\pi^2 n(\b{r}))^{1/3}$, $\hat{\b{R}}=\b{R}/R$, $z=2k_F R$ and
\begin{equation}
\label{}
J(z) = 72 [4+z^2 -(4-z^2)\cos z -4z \sin z ]/z^6,
\end{equation}
\begin{equation}
\label{}
L(z) = 9 (2 - 2 \cos z  -z\sin z)/z^3,
\end{equation}
\begin{equation}
\label{}
M(z) = 9 (  -z \cos z + \sin z)/(16z),
\end{equation}
\begin{equation}
\label{}
N(z) = 3 [8  - (8-4z^2)\cos z -(8z-z^3) \sin z ]/(16 z^4).
\end{equation}
The spherical average of the GEA exchange hole in term of the reduced gradient $s=|\nabla n|/(2 k_F n)$ and the reduced Laplacian $q=\nabla^2 n/((2 k_F)^2 n)$ writes
\begin{widetext}
\begin{equation}
n_x^{\text{GEA}}(k_F,s,q,z) = \frac{1}{4\pi} \int n_x^{\text{GEA}}(\b{r},\b{r}+\b{R}) d\Omega_{\b{R}} = n_x^{\text{unif}}(k_F,z) + n_x^{\text{grad}}(k_F,z) s^2 + n_x^{\text{lap}}(k_F,z) q,
\label{}
\end{equation}
with the exchange hole of the uniform electron gas
\begin{equation}
\label{}
n_x^{\text{unif}}(k_F,z) = \frac{24 k_F^3 [z \cos(z/2) -2 \sin(z/2)]^2}{\pi^2 z^6},
\end{equation}
the gradient coefficient
\begin{equation}
\label{}
n_x^{\text{grad}}(k_F,z) = \frac{k_F^3 [ -4(18+z^2) + (72-32z^2 +3z^4) \cos z +2 z (36-5z^2) \sin z]}{54 \pi^2 z^4},
\end{equation}
and the Laplacian coefficient
\begin{equation}
\label{}
n_x^{\text{lap}}(k_F,z) = \frac{k_F^3 [ 12(12+z^2) + 12(-12+5z^2) \cos z +12 z (-12+z^2) \sin z]}{54 \pi^2 z^4}.
\end{equation}
\end{widetext}

The exchange energy density is
\begin{eqnarray}
\lefteqn{\bar{\varepsilon}_{x}^{\text{GEA-Lap}}(n,s,q,\mut) = \frac{1}{2} \int n_x^{\text{GEA}}(\b{r},\b{r}+\b{R}) \bar{v}_{ee}^{\mu}(R) d\b{R}}
\nonumber\\
 && = \frac{1}{2 (2 k_F)^3} \int_{0}^{\infty} n_x^{\text{GEA}}(k_F,s,q,z) \bar{v}_{ee}^{\mu}(\frac{z}{2 k_F}) 4 \pi^2 z^2 dz,
\nonumber\\
\label{epsxintoverz}
\end{eqnarray}
where $\mut=\mu/(2 k_F)$. For convenience, $\bar{\varepsilon}_{x}^{\text{GEA-Lap}}(n,s,q,\mut)$ can be expressed by
\begin{equation}
\bar{\varepsilon}_{x}^{\text{GEA-Lap}}(n,s,q,\mut) = \bar{\varepsilon}_{x}^{\text{unif}}(n,\mut) \left[ 1 + a(\mut) s^2 + d(\mut) q \right],
\label{GEA-Lap}
\end{equation}
where $\bar{\varepsilon}_{x}^{\text{unif}}(n,\mut)$ has been given in Refs.~\onlinecite{Sav-INC-96,TouSavFla-IJQC-XX}. (Note that, for the \textit{erfgau} interaction, in Ref.~\onlinecite{TouSavFla-IJQC-XX} the constant $c$ of Eq.~(\ref{veeerfgau}) was absorbed into the parameter $\mu$). The integral in Eq.~(\ref{epsxintoverz}) diverges for $\mu=0$ due to the long-range character of the Coulomb interaction and thus $\bar{\varepsilon}_{x}^{\text{GEA-Lap}}(n,s,q,\mut=0)$ is not defined. (For a recent discussion on the non-analyticity of the inhomogeneous exchange energy density with respect to the electron density, see Refs.~\onlinecite{ArmMat-PRB-02,ArmMat-PRB-03}). Nevertheless, for finite $\mu$, the integral does exist and the gradient and Laplacian coefficients, $a(\mut)$ and $d(\mut)$, can be computed numerically.

To eliminate the divergence at $\mu=0$, one can perform an integration by parts of $n_x^{\text{GEA}}(k_F,s,q,z)$ over $\b{r}$ in Eq.~(\ref{Ex}) and define a new GEA exchange hole in term of the reduced gradient $s$ only
\begin{equation}
\label{}
\tilde{n}_x^{\text{GEA}}(k_F,s,z) = n_x^{\text{unif}}(k_F,z) + \tilde{n}_x^{\text{grad}}(k_F,z) s^2,
\end{equation}
with the associated gradient coefficient
\begin{widetext}
\begin{equation}
\label{}
\tilde{n}_x^{\text{grad}}(k_F,z) = - \frac{k_F^3 [ -72 + (72 -36 z^2 +z^4) \cos z-2 z (-36+5 z^2) \sin z]}{54 \pi^2 z^4},
\end{equation}
\end{widetext}
and the corresponding exchange energy density
\begin{eqnarray}
\lefteqn{\bar{\varepsilon}_{x}^{\text{GEA}}(n,s,\mut) = }
\nonumber\\
 && \frac{1}{2 (2 k_F)^3} \int_{0}^{\infty} \tilde{n}_x^{\text{GEA}}(k_F,s,z) \bar{v}_{ee}^{\mu}(\frac{z}{2 k_F}) 4 \pi^2 z^2 dz.
\nonumber\\
\label{epsxintoverz2}
\end{eqnarray}
Similarly to Eq.~(\ref{GEA-Lap}), $\bar{\varepsilon}_{x}^{\text{GEA}}(n,s,\mut)$ is expressed as
\begin{equation}
\label{}
\bar{\varepsilon}_{x}^{\text{GEA}}(n,s,\mut) = \bar{\varepsilon}_{x}^{\text{unif}}(n,\mut) \left[ 1 + b(\mut) s^2 \right].
\end{equation}
The integral in Eq.~(\ref{epsxintoverz2}) is now defined for all $\mu's$ and can be done analytically for both the \textit{erf} and \textit{erfgau} modified interactions. For the gradient coefficient $b(\mut)$ with the \textit{erf} interaction, we obtain
\begin{equation}
\label{berf}
b_{\erf}(\mut) = \frac{-c_1 +c_2 e^{1/(4\mut^2)}}{c_3 +54 c_4 e^{1/(4\mut^2)}},
\end{equation}
and, with the \textit{erfgau} interaction,
\begin{equation}
\label{berfgau}
b_{\erfgau}(\mut) = \frac{-3\sqrt{3}c_5 +2\nu^2 c_1 e^{1/(2\nu^2)} + c_6 e^{3/(4\nu^2)}}{c_7 -2\nu^2 c_3 e^{1/(2\nu^2)} +12 \nu^2 (-9 c_4 + c_8) e^{3/(4\nu^2)} },
\end{equation}
with $c_1=1 +22\nu^2 +144\nu^4$, $c_2=2\nu^2(-7+72\nu^2)$, $c_3=-864\nu^4(-1+2\nu^2)$, $c_4=\nu^2[-3 -24\nu^2 +32\nu^2 +8\nu\sqrt{\pi}\erf(1/(2\nu))]$, $c_5=3+18\nu^2+48\nu^4+64\nu^6$, $c_6=4\nu^4(7-72\nu^2+48\sqrt{3}\nu^2)$, $c_7=-192\sqrt{3}\nu^6(-3+8\nu^2)$, $c_8=8\nu^3[-18\sqrt{3}\nu+16\sqrt{3}\nu^3+9\sqrt{\pi}\erf(\sqrt{3}/(2\nu))]$, $\nu=\mut/c$ and $c = \left( 1+6\sqrt{3}\right)^{1/2} \approx 3.375$. For $\mut=0$ for both interactions, $b(\mut)$ reduces to Sham's coefficient~\cite{Sha-INC-71,Per-PRL-85} of the standard exchange GEA
\begin{equation}
\label{}
b(\mut=0) = \frac{7}{81}.
\end{equation}

The approximations $\bar{\varepsilon}_{x}^{\text{GEA-Lap}}(n,s,q,\mut)$ and $\bar{\varepsilon}_{x}^{\text{GEA}}(n,s,\mut)$ will be referred to as $\mu$-GEA-Lap and $\mu$-GEA, respectively. They are compared in Sec.~\ref{sec:gea} for the case of the Be atom with the \textit{erfgau} interaction (Fig.~\ref{fig:epsx-be-erfgau-mu0.21-gea}). Because of its greatest simplicity, $\bar{\varepsilon}_{x}^{\text{GEA}}(n,s,\mut)$ will be used in practice.

\begin{figure}
\includegraphics[scale=0.75]{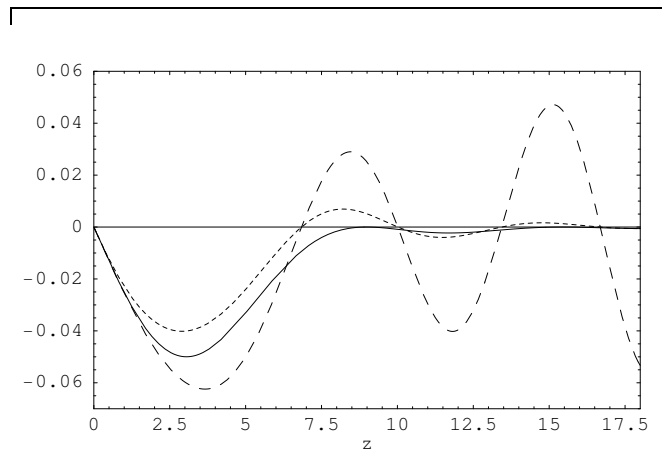}
\caption{z-integrand of the short-range $\mu$-GEA exchange energy density $\bar{\varepsilon}_{x}^{\text{GEA}}(n,s,\mut)$ (Eq.~\ref{epsxintoverz2}) with the \textit{erf} interaction and with $s=0$ and $\mu=0$ (full curve), $s=1$ and $\mu=0$ (long-dashed curve) and $s=1$ and $\mu=0.2$ (short-dashed curve), all for $k_F=1$.
}
\label{fig:epsintegrand-erf-gea}
\end{figure}

To see the effect the short-range interaction $\bar{v}_{ee}^{\mu}(R)$, the $z$-integrand of the $\mu$-GEA exchange energy density (Eq.~\ref{epsxintoverz2}) has been plotted with respect to $z$ in Fig.~\ref{fig:epsintegrand-erf-gea} in the case of the \textit{erf} interaction. One sees that the spurious strong oscillations of the GEA exchange hole at large interelectronic distances are efficiently cut off at finite $\mu$ by the short-range interaction. In other words, only the short-range part of the GEA exchange hole is used in the short-range exchange functional.

\begin{figure}
\includegraphics[scale=0.75]{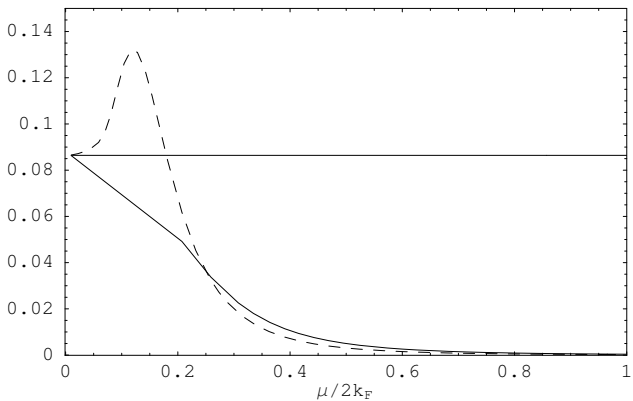}
\caption{Ratio of the gradient contribution of $\mu$-GEA short-range exchange energy density to the LDA exchange energy density with Coulomb interaction $\bar{\varepsilon}_{x}^{\text{unif}}(n,\mut)b(\mut)/\varepsilon_{x}^{\text{unif}}(n)$ with respect to $\mu/(2 k_F)$ for the \textit{erf} (full curve) and \textit{erfgau} (dashed curve) interactions. The horizontal line corresponds to the Coulomb case ($\mu=0$) where $b(\mu=0)=7/81$.
}
\label{fig:epsxmubmu-erferfgau}
\end{figure}

In order to appreciate the contribution of the gradient term of the $\mu$-GEA with respect to the interaction parameter, we have represented $\bar{\varepsilon}_{x}^{\text{unif}}(n,\mut)b(\mut)/\varepsilon_{x}^{\text{unif}}(n)$ in Fig.~\ref{fig:epsxmubmu-erferfgau} for the \textit{erf} and \textit{erfgau} interactions. Not surprisingly, the gradient correction is decreased for a short-range interaction. Notice however that for small $\mut$, the gradient correction for the \textit{erfgau} is increased compared to the Coulomb case $\mut=0$.

\section{Short-range PBE functional}
\label{app:PBE}

In the context of Kohn-Sham DFT, Perdew, Burke and Ernzerhof~\cite{PerBurErn-PRL-96} have proposed a gradient-corrected functional where the parameters are determined so as to satisfy theoretical conditions. We propose here to generalize the PBE functional along the \textit{erf} or \textit{erfgau} adiabatic connection. We note that for a different purpose Heyd, Scuseria and Ernzerhof~\cite{HeyScuErn-JCP-03} has already derived a short-range version of the exchange PBE functional corresponding to the \textit{erf} interaction by using a model of the PBE exchange hole~\cite{ErnPer-JCP-98}. (The obtained functional is called $\omega$PBE where $\omega$ corresponds to our parameter $\mu$). We shall not follow this procedure but come back instead to the original formulation of the PBE functional.

\begin{figure}
\includegraphics[scale=0.75]{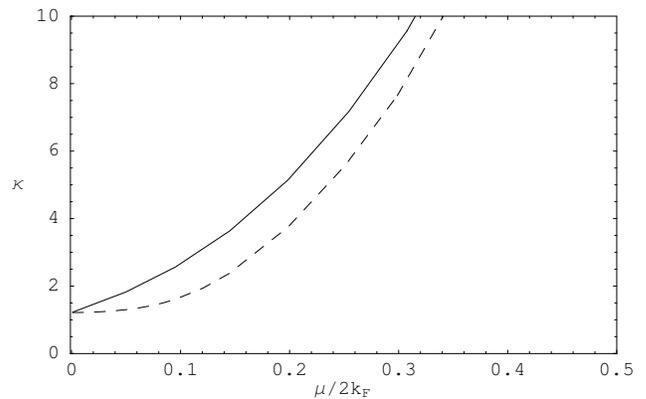}
\caption{Coefficients $\kappa$ of the $\mu$-PBE exchange functional with respect to $\mu/(2 k_F)$ for the \textit{erf} (full curve) and \textit{erfgau} (dashed curve) interactions.
}
\label{fig:kappamu-erferfgau}
\end{figure}

We thus take for the short-range exchange energy functional the PBE form
\begin{equation}
\bar{E}_{x}^{\mu} =  \int d\b{r} n(\b{r}) \bar{\varepsilon}_{x}^{\text{unif}}(n(\b{r}),\mut(\b{r})) F_x(s(\b{r}),\mut(\b{r})),
\label{ExPBE}
\end{equation}
where we have introduced a dependence on the reduced interaction parameter $\mut=\mu/(2 k_F)$. In Eq.~(\ref{ExPBE}), $s=|\nabla n|/(2 k_F n)$ is a reduced density gradient, $\bar{\varepsilon}_{x}^{\text{unif}}(n,\mut)$ is the exchange energy per particle of a uniform electron gas with short-range interaction (see Refs.~\onlinecite{Sav-INC-96,TouSavFla-IJQC-XX}), and $F_x^{\mu}(s,\mut)$ is the enhancement factor
\begin{equation}
\label{}
F_x(s,\mut) = 1 +\kappa(\mut) -\frac{\kappa(\mut)}{1+b(\mut) s^2/\kappa(\mut)}.
\end{equation}
Note that this form had also be proposed by Becke~\cite{Bec-JCP-86}. The constant $b(\mut)$ is fixed by imposing the correct gradient expansion of $F_x(s,\mut)$ as $s \to 0$
\begin{equation}
\label{}
F_x(s,\mut) \approx 1 + b(\mut) s^2 +\cdots,
\end{equation}
while $\kappa(\mut)$ is connected to the rapidly varying limit ($s \to \infty$)
\begin{equation}
\label{}
F_x(s,\mut) \to 1 + \kappa(\mut).
\end{equation}
For $b(\mut)$ we take the coefficient arising from the gradient expansion of the short-range exchange energy with \textit{erf} or \textit{erfgau} interaction (see Appendix~\ref{app:geax}, Eqs.~\ref{berf} and~\ref{berfgau}). $\kappa(\mut)$ is determined by imposing the Lieb-Oxford bound~\cite{LieOxf-IJQC-81} which still holds for the short-range exchange functional $\bar{E}_{x}^{\mu}$
\begin{equation}
\label{}
\bar{E}_{x}^{\mu} \geq E_{x} \geq -C \int n(\b{r})^{4/3} d\b{r},
\end{equation}
since $\bar{E}_{x}^{\mu} = E_{x} - E_{x}^{\mu}$ and the long-range exchange energy $E_{x}^{\mu}$ is always negative. The constant $C$, for which Lieb and Oxford originally found $1.6787$, has recently be improved by Chan and Handy to a value $C=1.6358$~\cite{ChaHan-PRA-99}. A sufficient (but not necessary) condition for this bound to be satisfied is
\begin{equation}
\label{fxLO}
F_{x}(s,\mut) \leq -C n^{1/3}/\bar{\varepsilon}_{x}^{\text{unif}}(n,\mut).
\end{equation}
We thus take $\kappa(\mut)=-C n^{1/3}/\bar{\varepsilon}_{x}^{\text{unif}}(n,\mut) -1$, the largest value insuring condition~(\ref{fxLO}). Fig.~\ref{fig:kappamu-erferfgau} shows $\kappa(\mut)$ for the \textit{erf} and \textit{erfgau} interaction. One sees that $\kappa(\mut)$ increases with
$\mut$, but the Lieb-Oxford bound is actually of no effect for large $\mut$ since the GEA gradient correction vanishes anyway and the enhancement factor reduces to $1$.

The short-range correlation energy is given by
\begin{equation}
\label{}
\bar{E}_{c}^{\mu} = \int d\b{r} n(\b{r}) [\bar{\varepsilon}_{c}^{\text{unif}}(r_s(\b{r}),\mu) + H(r_s(\b{r}),t(\b{r}),\mu)],
\end{equation}
with the reduced density gradient $t=|\nabla n|/(2 k_s n)$. We imposed the same conditions on the gradient correction $H(r_s,t,\mu)$ as PBE.

(a) The second-order gradient expansion in the slowly varying limit ($t \to 0$) is
\begin{equation}
\label{}
H(r_s,t,\mu) \approx \beta(r_s,\mu) t^2 +\cdots.
\end{equation}
where the coefficient $\beta(r_s,\mu)$ is estimated from the exchange gradient expansion according to Eq.~(\ref{betamu}).

(b) The correlation is set to zero in the rapidly varying limit ($t \to \infty$), thus
\begin{equation}
\label{}
H(r_s,t,\mu) \to -\bar{\varepsilon}_{c}^{\text{unif}}(r_s,\mu).
\end{equation}

(c) Under uniform scaling to the high-density limit, i.e. $n(\b{r}) \to \lambda^3 n(\lambda \b{r})$, $r_s \to \lambda^{-1}r_s$, $t \to \lambda^{1/2} t$  and $\lambda \to \infty$, the correlation energy scales to a constant. This well-known property for the Coulomb interaction case~\cite{Lev-PRA-91} is also valid for the short-range correlation functional. Thus, $H(r_s,t,\mu)$ must cancel the logarithmic divergence of the local short-range correlation energy $\bar{\varepsilon}_{c}^{\text{unif}}(\lambda^{-1} r_s,\mu) \approx \gamma \ln(\lambda^{-1}) +\cdots$ with $\gamma \approx 0.031091$, appearing as in the Coulomb case. Thus, as $\lambda \to \infty$,
\begin{equation}
\label{}
H(\lambda^{-1} r_s,\lambda^{1/2} t,\mu) \approx \gamma \ln (\lambda) +\cdots.
\end{equation}
We then take the same ansatz than PBE
\begin{equation}
\label{}
H(r_s,t,\mu) = \gamma \ln \left[ 1 + \frac{\beta(r_s,\mu)t^2}{\gamma} \left( \frac{1 + At^2}{1 + At^2 + A^2 t^4} \right) \right],
\end{equation}
with
\begin{equation}
\label{}
A=\frac{\beta(r_s,\mu)}{\gamma} \frac{1}{e^{-\bar{\varepsilon}_{c}^{\text{unif}}(r_s,\mu)/\gamma }-1}.
\end{equation}

Note that in the original PBE functional, the second-order gradient coefficient for correlation was imposed in the high-density limit, $\beta(r_s \to 0,\mu=0)=\beta=0.066725$ and the second-order gradient coefficient for exchange was chosen so that to exactly cancel the correlation gradient expansion: $b(\mu=0)=\beta(\pi^2/3)=0.21951$. On the contrary, we have used the exact exchange gradient expansion for finite $\mu$ and estimate the correlation gradient expansion from it. Therefore, our $\mu$-PBE functional does not reduce to the original PBE functional for $\mu=0$. 

%BIBLIOGRAPHY---------------------------------------------
\bibliographystyle{/home/toulouse/tex/biblio/apsrev}
\bibliography{/home/toulouse/tex/biblio/biblio}

%FIGURES--------------------------------------------------

\end{document}